\documentclass[aps,prl,twocolumn,showpacs,superscriptaddress]{revtex4-1}
\usepackage{amsfonts}

\usepackage{graphicx}
\usepackage{subfigure}
\usepackage{braket}
\usepackage{amsmath}
\usepackage{bm}

\begin{document}

\title{Quantum Dynamics of the Avian Compass}
\author{Zachary~B.~Walters}
\affiliation{Max Planck Institute for Physics of Complex Systems, N\"othnitzer
Strasse 38, D-01187 Dresden, Germany}
\date{\today}

\begin{abstract}
  The ability of migratory birds to orient relative to the Earth's magnetic
  field is believed to involve a coherent superposition of two spin states of
  a radical electron pair.  However, the mechanism by which this coherence can
  be maintained in the face of strong interactions with the cellular
  environment has remained unclear.  This Letter addresses the problem of
  decoherence between two electron spins due to hyperfine interaction with a
  bath of spin 1/2 nuclei.  Dynamics of the radical pair density matrix are
  derived and shown to yield a simple mechanism for sensing magnetic field
  orientation.  Rates of dephasing and decoherence are calculated {\em ab
    initio} and found to yield millisecond coherence times, consistent with
  behavioral experiments.
\end{abstract}

\maketitle

The ability of a migratory bird to orient itself relative to the Earth's
magnetic field is at once a familiar feature of everyday life and a puzzling
problem of quantum mechanics.  That birds have this ability is well
established by a long series of behavioral experiments.  However, the precise
mechanism by which an organism may sense the orientation of the weak
geomagnetic field remains unclear and theoretically problematic.

Although commonly referred to as the ``avian compass,'' an ability to
sense the local magnetic field orientation has been observed in every
major group of vertebrates, as well as crustaceans, insects, and a
species of mollusc \cite{wiltschko2005magnetic,muheim2008light}.  For
the majority of species, the primary compass mechanism appears to be
light-activated, with a few exceptions such as the sea turtle or the
subterranean mole rat\cite{muheim2008light}.  In addition to a
light-activated compass located in the eye, migratory birds are
believed to possess a separate mechanism involving magnetite, with
possible receptors identified in the beak\cite{wiltschko2010role}, the
middle ear\cite{wu2011magnetoreception} and the brain
stem\cite{wu2012neural}, although the existence of a receptor in the
beak has been challenged in a recent study\cite{treiber2012clusters}.
This paper addresses the light-activated mechanism, which in addition
to being widespread is also well studied by a long series of
behavioral experiments, reviewed in
\cite{muheim2008light,wiltschko2011mechanism,ritz2011quantum,
  johnsen2005physics,able1995orientation}.

The basic parameters of the compass mechanism may be probed by confining a
bird in a conical cage during its preferred migration period
\cite{beck1983orientation}.  The restless nocturnal hopping behavior, or
Zugunruhe, will tend to orient in the preferred migration direction, and the
effects of environmental parameters can be judged by whether they affect the
bird's ability to orient.  Such experiments have established that the compass
is light activated, with an abrupt cutoff between wavelengths 560.5 and 567.5
nm \cite{muheim2002magnetic}, and that birds are sensitive to the orientation
of magnetic field lines but not their polarity -- they cannot distinguish
magnetic north from south\cite{wiltschko1972magnetic}.  Provocatively, a
recent experiment has found that an oscillatory magnetic field oriented
transverse to the static field can cause disorientation when it is narrowly
tuned to the Larmor frequency for an electron in the static field to flip its
spin.  On resonance, an oscillatory field strength of 15 nT (Rabi frequency
$\Omega_{\text{Rabi}}=1320$ Hz) is sufficient to cause
disorientation\cite{ritz2009magnetic,ritz2004resonance}.

Qualitatively, such experiments are well explained by a ``radical pair'' model
of the avian compass\cite{schulten1978biomagnetic,ritz2000model}, also known as
the Ritz model.  Here, an asymmetry $\Delta \mu=\mu_{1}-\mu_{2}$ in the
coupling of the magnetic field to the two electrons, arising due to chemical or
physical properties of the receptor, allow the magnetic field to drive coherent
oscillations between the $\Ket{s,m_{s}}=\Ket{0,0}$ singlet state $\Ket{s}$ and
the $\Ket{s,m_{s}}=\Ket{1,0}$ triplet state $\Ket{t}$ of an electron radical
pair formed by absorption of a photon.  If the singlet and triplet states react
to form distinguishable byproducts, or can be otherwise
distinguished\cite{stoneham2012new}, monitoring the ratio of the byproducts
probes the time spent in each state, and thus the oscillation frequency.  Use
of a radical pair is a common denominator in a wide variety of biological
processes sensitive to magnetic fields, recently reviewed in
\cite{buchachenko2014magnetic}.

The radical pair model gives an excellent phenomenological description of the
avian compass, and predicts disorientation by an on-resonance oscillatory
field.  However, it remains theoretically problematic, requiring that
coherence be maintained between different spin states for very long times
despite the presence of an environment which is very hostile to this.  As
observed in \cite{gauger2011sustained}, the slow spin flip time
($\pi/\Omega_{\text{Rabi}}=3$ms) implies that the process it disrupts must be
slower still. \cite{gauger2011sustained} and
\cite{bandyopadhyay2012quantum,bandyopadhyaycomment,bandyopadhyaycommentreply} 
use similar methods to infer coherence times of $10^{-5}-10^{-4}$s.  However,
the proteins and water molecules present in a cellular environment possess
large numbers of hydrogen nuclei, each of which interact with the radical pair
via the hyperfine interaction.  Somehow, the necessary quantum information
must survive such interactions long enough to give a biologically useful
signal.

Previous work considering the radical pair compass in the presence of
decoherence includes \cite{kominis2009quantum, jones2010spin,
  kominis2011radical, dellis2011quantum}, treating effects of rapid singlet
and triplet reaction rates on the evolution of the density matrix.
Decoherence due to hyperfine interactions has been treated in terms of an
effective magnetic field in \cite{kavokin2009puzzle}, while
\cite{cai2010quantum,tiersch2012decoherence,cai2012quantum} consider a radical
pair interacting with a small number of nuclei.  The related problem of
decoherence in a singlet/triplet quantum dot has been treated in
\cite{johnson2005triplet,petta2005coherent}.


This paper gives an analytic treatment of the preservation and decay of
coherence for a radical pair interacting with a bath of spin 1/2 nuclei.  A
long lived component of the quantum information is identified, and shown to
yield a simple and robust compass mechanism.  Design considerations for an
efficient compass are identified, and the coherence lifetime is shown to be
consistent with lifetimes inferred from behavioral experiments.  Atomic units
are used throughout.

To maximize readability, the text of this paper is split into two parts.  The
body of the paper addresses the classic Ritz model of the radical pair compass
with the addition of decoherence terms which are included as Lindblad
superoperators.  Because the Ritz model addresses only dynamics within the two
state $m_{z}=0$ subspace of the radical pair, only these two states are
included.  The Ritz model assumes that the two electrons experience a slightly
different Zeeman coupling to the local magnetic field; as the receptor or
receptors involved in the avian compass are currently unknown, this paper
incorporates this assumption without proof.  Both the eigencomponents of
dephasing induced by the hyperfine interaction and their rates of decay are
derived in the technical appendix, which includes all four states of the
radical pair, plus two states of the nuclear spin.  The asymmetric Zeeman
coupling assumed by the Ritz model is not included in the derivation of
dephasing rates, but could readily be added using the same approach.

The evolution of the reduced density matrix $\rho$ for a radical pair
interacting with a Markovian bath is given by the Lindblad master equation
\begin{equation}
\frac{\partial}{\partial t}\rho=i[H^{rp}_{0},\rho]+
\sum_{\kappa}
\Gamma_{\kappa}\mathcal{L}_{\kappa}[\rho],
\label{eq:lindblad}
\end{equation}
Here 
\begin{equation}
H^{rp}_{0}=\bar{\mu}(\vec{s}_{1}+\vec{s}_{2})\cdot \vec{B}+ \Delta
\mu(\vec{s}_{1}-\vec{s}_{2})\cdot \vec{B}
\label{eq:H0rp}
\end{equation}
is the Zeeman Hamiltonian, and
\begin{equation}
\mathcal{L}_{\kappa}[\rho]=-\rho L_{\kappa}^{\dag} L_{\kappa}-
L_{\kappa}^{\dag} L_{\kappa} \rho+2 L_{\kappa}\rho L_{\kappa}^{\dag}
\end{equation}
is the
Lindblad superoperator corresponding to projection operator
$L_{\kappa}=\Ket{\kappa}\Bra{\kappa}$.  
The difference $\Delta \mu$ in the magnetic
susceptibilities of the two electrons  is assumed to arise due to short range
interactions with the receptor
molecule\cite{cai2011quantum,cai2012quantum,cai2012sensitive}; as the receptor
or receptors involved in the avian compass are as yet
unknown,\cite{liedvogel2010cryptochromes, hogben2009possible,
niessner2011avian}, this paper simply assumes a value of $\Delta \mu \approx1$
without derivation.  

If all Lindblad operators in Eq. \ref{eq:lindblad} were zero, the equation
would recover the Ritz model of the avian compass, in which decoherence is
ignored.  Following the Ritz model, theoretical treatments of the avian compass
have frequently assumed that rates of decay are slow relative to the dynamics
induced by the Zeeman Hamiltonian.  However, as shall be shown here, the limit
of rapid dephasing also allows for an efficient compass, with a reaction product signal which is relatively easy and unambiguous to interpret.

As derived in the appendix, hyperfine interactions between the
electronic spins of the radical pair and the nuclear spins of atoms in the
surrounding environment -- most plentifully, hydrogen atoms in the surrounding
water molecules -- causes a loss of coherence between spin states of the
radical pair.  Although both the singlet and the triplet state have total spin
$m_{s}=0$, the hyperfine interaction couples the triplet state to other triplet
states with $m_{s}=\pm 1$, while the singlet state is not coupled to any other
states.  Because of this, the spin state of the radical pair becomes entangled
with the unobserved spin states of the bath nuclei, and coherence between
different states of the radical pair decays with time.

At the same time that coherence decays due to hyperfine interaction with the
bath, it is being created by the normal Hamiltonian evolution which arises when
two states are connected by a matrix element.  The evolution of the density
matrix includes both effects, and in the limit that the decay rate is very
large, they become very closely balanced against each other for a particular
component of the density matrix, which accordingly decays very slowly.  Because
the other components of the density matrix decay rapidly, the density matrix
describing the radical pair soon evolves to consist of only the long lived
component.  As will be seen, the rate of decay for this long lived component
gives all the information necessary for an efficient chemical compass.  Because
this decay manifests itself as a transfer of population from the singlet to the
triplet state, it is well suited to detection by spin selective chemical
reactions which create different sets of byproducts depending on whether the
radical pair is in the triplet or singlet state.

As derived in the appendix, the rates of decay relevant to the avian compass
are given by two parameters, which can be found analytically.  In Eq.
\ref{eq:lindblad}, $\Gamma_{\Ket{s}}=\Gamma_{\Ket{t}}=\bar{\Gamma}/2$ and
$\Gamma_{\Ket{\uparrow\downarrow}}=\Gamma_{\Ket{\downarrow\uparrow}}=\Delta
\Gamma/2$, where $\bar{\Gamma}$ is large for moderate field strengths and
$\Delta \Gamma$ is zero for some orbital symmetries.  Mapping the density
matrix to a Bloch sphere according to
$(\rho_{ss}-\rho_{tt})\rightarrow(\rho_{01}+\rho_{10})=x \sigma_{x}$,
$(\rho_{st}-\rho_{ts})\rightarrow(\rho_{10}-\rho_{01})=iy\sigma_{y}$, and
$(\rho_{st}+\rho_{ts})\rightarrow(\rho_{00}-\rho_{11})=z\sigma_{z}$, where
$\sigma_{x,y,z}$ are Pauli matrices, it can be seen that a Lindblad operator
corresponding to projecting the Bloch vector in one direction causes decay of
vectors perpindicular to that direction, while the difference in magnetic
susceptibilities $\Delta \mu$ causes the Bloch vector to precess about the z
axis when a magnetic field is present, thereby creating coherence between $\ket{s}$ and
$\ket{t}$. In the Bloch sphere picture, a quantum mechanically pure state
corresponds to a vector with length 1, while a completely incoherent state
corresponds to a vector of length 0.  Hamiltonian evolution rotates the Bloch
vector about some axis, while dephasing causes some components of the vector to
decay.  In the work that follows, it is useful to draw a distinction between
the rate of {\em dephasing} -- the rate at which these components would decay
if there were no Hamiltonian evolution, and the rate of {\em decoherence} --
the rate at which the length of the Bloch vector decays when both dephasing and
Hamiltonian evolution are taken into account.  As will be seen, a large rate of
dephasing may paradoxically lead to a small rate of decoherence.  This is the
quantum mechanical version of Zeno's paradox, and is appropriately known as the
quantum Zeno effect\cite{home1997conceptual,kominis2009quantum,dellis2011quantum}.

The evolution of the density matrix components can be found analytically by
calculating the evolution due to $H$ and $\bar{\Gamma}$ in a basis where
singlet and triplet states form the $\pm z$ axes in the Bloch sphere, then
transforming to a basis where  $\ket{\uparrow\downarrow}$ and
$\ket{\downarrow\uparrow}$ make up the $z$ axis to include the effects of
$\Delta \Gamma$.  As in \cite{walters2011quantum}, differential equations for
the density matrix components due to $H$ and $\bar{\Gamma}$ are given by
\begin{equation} \begin{split}
	&\frac{d^{2}}{dt^{2}}(\rho_{ss}-\rho_{tt})+\bar{\Gamma}
	\frac{d}{dt}(\rho_{ss}-\rho_{tt})+ (B_{z}\Delta
	\mu)^{2}(\rho_{ss}-\rho_{tt})=0 \\
	&\frac{d^{2}}{dt^{2}}(\rho_{ts}-\rho_{st})+\bar{\Gamma}
	\frac{d}{dt}(\rho_{ts}-\rho_{st})+ (B_{z}\Delta
	\mu)^{2}(\rho_{ts}-\rho_{st})=0 \\
	&\frac{d}{dt}(\rho_{ts}+\rho_{st})=-\bar{\Gamma}(\rho_{ts} +\rho_{st})
	\\ &\frac{d}{dt}(\rho_{ss}+\rho_{tt})=0, \label{eq:blochDEs}
\end{split} \end{equation}
where $\rho_{ss}$  gives the population of singlet states, $p_{st}$ is a coherence term between singlet and triplet, and so on.

\begin{figure}
\begin{center}
\includegraphics[width=\columnwidth]{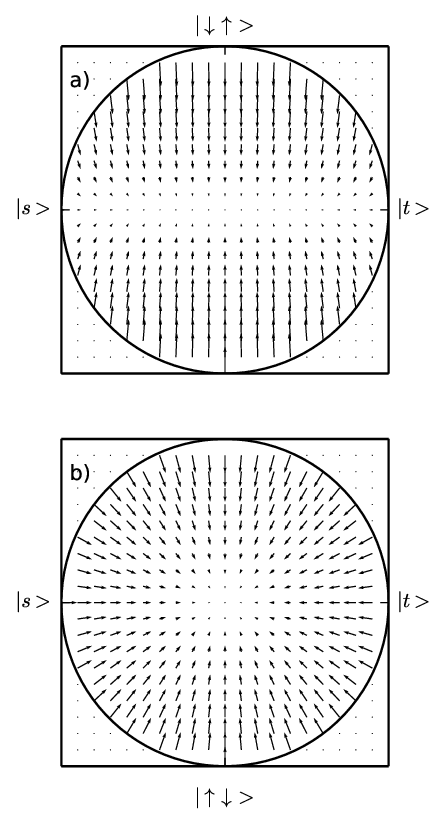}
\end{center}
\caption{ Decoherence of the Bloch vector,
  $\dot{\vec{V}}_{B}(\vec{x})$, in the $xz$ plane for an efficient and
  an inefficient compass molecule.  The $z$ component decays as
  $e^{-\bar{\Gamma}t}$, the $x$ as $e^{-\lambda_{+}t}$.  a)
  $\frac{\bar{\Gamma}}{B \Delta \mu}=8.5$, $\frac{\Delta \Gamma}{B
    \Delta \mu}=6\times 10^{-4}$,
  contrast=$0.99$. b)$\frac{\bar{\Gamma}}{B \Delta \mu}=8.5$,
  $\frac{\Delta \Gamma}{B \Delta \mu}=5.8$, contrast=$0.01$.  For both
  plots, a small value of $\bar{\Gamma}$ has been used to accentuate
  the decay of the $x$ component.  }
\label{fig:coherencedecay}
\end{figure}

In Eq. \ref{eq:blochDEs},
$(\rho_{ss}-\rho_{tt})$ and $(\rho_{ts}-\rho_{st})$ behave as damped harmonic
oscillators, with time dependence $P(t)=Ae^{\lambda_{+}t}+Be^{\lambda_{-}t}$,
where 
\begin{equation}
	\lambda_{\pm}=\frac{-\bar{\Gamma} \pm
	\sqrt{\bar{\Gamma}^{2}-4(B_{z}\Delta \mu)^{2}}}{2}.
	\label{eq:lambdapm}
\end{equation}
In the limit that $\bar{\Gamma}>>|2 B_{z} \Delta \mu|$, the system is strongly
overdamped and the coherence terms will decay much more slowly than the base
rate of dephasing, $\bar{\Gamma}$.  In the Bloch sphere picture, the $z$
component of the Bloch vector decays rapidly, while the $x$ and $y$ components
decay slowly.  It is this slow loss of coherence, shown in Figure
\ref{fig:coherencedecay}, which allows for a biologically useful signal.


Although arising from a different source, these dynamics are similar to the
quantum Zeno regime treated in \cite{kominis2009quantum,dellis2011quantum},
where fast singlet or triplet reaction rates take the place of rapid
dephasing, and to \cite{walters2011quantum}, where the long lived coherences
occur in photosynthetic molecules.  Because the $z$ component of the Bloch
vector decays rapidly, the symmetry group of the long lived information is
$U(1)$ rather than $SU(2)$.

The value of $\bar{\Gamma}=\frac{5}{3} B \Delta \mu N_{\text{sphere}}$
derived in the appendix can be found for a cellular
environment by assuming a density of hydrogen nuclei equal to that of
liquid water.  For B=50 $\mu$T, $N_{\text{sphere}}=3300$ is the number
of nuclei within radius $r_{0}=43$ bohr, at which the hyperfine
interaction equals the Zeeman interaction in magnitude.  The dynamics
are thus strongly overdamped, with a dephasing lifetime
$\bar{\Gamma}^{-1}=43$ ps and a coherence lifetime
$\tau=\lambda_{+}^{-1}=1.3$ ms, somewhat longer than the
$10^{-4}$s-$10^{-6}$s inferred in
\cite{gauger2011sustained,bandyopadhyay2012quantum}.

The dynamics of the overdamped radical pair model depart in an essential way
from those of the Ritz model, or from a model in which dephasing is present but
weak. Because the decay of the Bloch vector is overdamped, it does not precess
about the z axis as in the original radical pair model.  Rather, a vector in
the equatorial plane is frozen in place and evolves only through decoherence.
For an initially pure singlet state, $\rho(0)=\left( \begin{array}{c c}1/2 &
	1/2 \\ 1/2 & 1/2 \end{array} \right)$ in the
$\Ket{\uparrow\downarrow}$,$\Ket{\downarrow\uparrow}$ basis, so that
$\dot{\rho}=-1/2\left( \begin{array}{c c}0 & \lambda_{+}+\Delta \Gamma \\
	\lambda_{+}+\Delta \Gamma & 0\end{array} \right)$, or $1/2\left(
	\begin{array}{c c}-\lambda_{+}-\Delta \Gamma & 0 \\ 0 &
		\lambda_{+}+\Delta \Gamma \end{array} \right)$ in the
		$\Ket{s}$, $\Ket{t}$
basis.  Loss of coherence thus manifests itself as a transfer of population
from singlet to triplet at a rate which varies as $B^{2}\cos^{2}{\theta}$.
Identical logic applies if the initial state is a triplet.   As this rate of
population transfer contains the necessary directional information, a chemical
compass requires only that the state which is not originally populated (here,
the triplet) have a reaction rate sufficiently large to prevent backwards
population transfer.  Population transfer due to $\Delta \Gamma$, which does
not depend on the orientation of the molecule, decreases the sensitivity of the
compass by decreasing contrast between orientations with a high rate of
transfer and orientations with a slow rate.  Assuming that the triplet reaction
rate is sufficiently high to prevent backwards population transfer, the ratio
of triplet to singlet byproducts is
\begin{equation}
  R_{ts}(\theta)=\frac{\lambda_{+}+\Delta \Gamma}{k_{s}}\approx\frac{|B  \Delta \mu\cos{\theta}
    |^{2}}{\bar{\Gamma} k_{s}}+\frac{\Delta \Gamma}{k_{s}},
\label{eq:Rts}
\end{equation}
where $k_{s}$ is the singlet reaction rate.
Note that a simple consequence of this model is that the chemical compass is insensitive to the difference between positive and negative values of $B \cos{\theta}$ -- ie, it is insensitive to the difference between magnetic North and South.  This is consistent with behavioral experiments, in which the inclination of field lines to the horizon, rather than their polarity, determines the preferred migratory direction.

While the identity of the avian compass receptor remains unknown, a
number of design considerations may be inferred from Eq. \ref{eq:Rts}
and from the dephasing dynamics derived in the appendix.


One such consideration relates to the mechanism of detecting
the formation of triplet states.  While the original radical pair
model proposed a spin sensitive chemical reaction, this is not an
essential feature of the model, and more recent papers
\cite{stoneham2012new} have proposed that physical detection of the
triplet states may be advantageous.  A possible mechanism for such
detection can be seen in Table \ref{table:Sbar_smallalpha} in the
appendix, which
shows that dephasing due to nuclei distant from the radical pair will
result in population transfer from state $\Ket{t}$ to states
$\Ket{t^{\pm}}$, with lifetime 33 ps.  As the $m_{s}=\pm1$ states have
nonzero magnetic moments, they are easily distinguishable from the
$m_{s}=0$ states by physical means.  Because equilibration between the
populations of states $\Ket{t}$, $\Ket{t^{+}}$ and $\Ket{t^{-}}$ is
rapid, detection of any triplet state will suffice for the purposes of
the compass mechanism.

Second, it can be seen that the sensitivity of the compass mechanism
depends greatly upon the form taken by the dephasing superoperators.
An upper limit for the sensitivity of the compass mechanism may be
found by considering the contrast between North/South and East/West
alignment
\begin{equation}
\text{contrast}=\frac{R_{ts}(0)-R_{ts}(\pi/2)}{R_{ts}(0)+R_{ts}(\pi/2)}=
\frac{(B_{0}\Delta \mu)^{2}}{2 \Delta \Gamma \bar{\Gamma}+(B_{0}\Delta
  \mu)^{2}}.
\label{eq:contrast}
\end{equation}
Here the contrast is independent of $k_{s}$ and $k_{t}$, depending
only upon the ratio of $\bar{\Gamma}\Delta\Gamma$ and
$(B\Delta\mu)^{2}$.  Figure \ref{fig:coherencedecay} illustrates the
decay of the Bloch vector for both an efficient (high contrast) and an
inefficient (low contrast) compass.  As $\bar{\Gamma}$ is large
relative to $B\Delta\mu$, it follows that an efficient compass
receptor must have $\Delta \Gamma$ small or zero.  

From table \ref{table:deltaS_smallalpha} in the appendix, it can be seen that
$\Delta \Gamma$ will be small only in the case that it is zero by symmetry.
Here the rate of dephasing $\frac{5\alpha \kappa}{6 \beta}$ due to
$\vec{I}\cdot\Delta \vec{S}$ for a nucleus far from the radical pair is
inversely proportional to $\tau_{\epsilon}=1/\kappa$, the rate of decay for
correlations in the environment.  Thus, it is likely that an efficient compass
will employ an excited state with cylindrical symmetry, which eliminates this
term.

\begin{figure}
\begin{center}
  \includegraphics[width=\columnwidth]{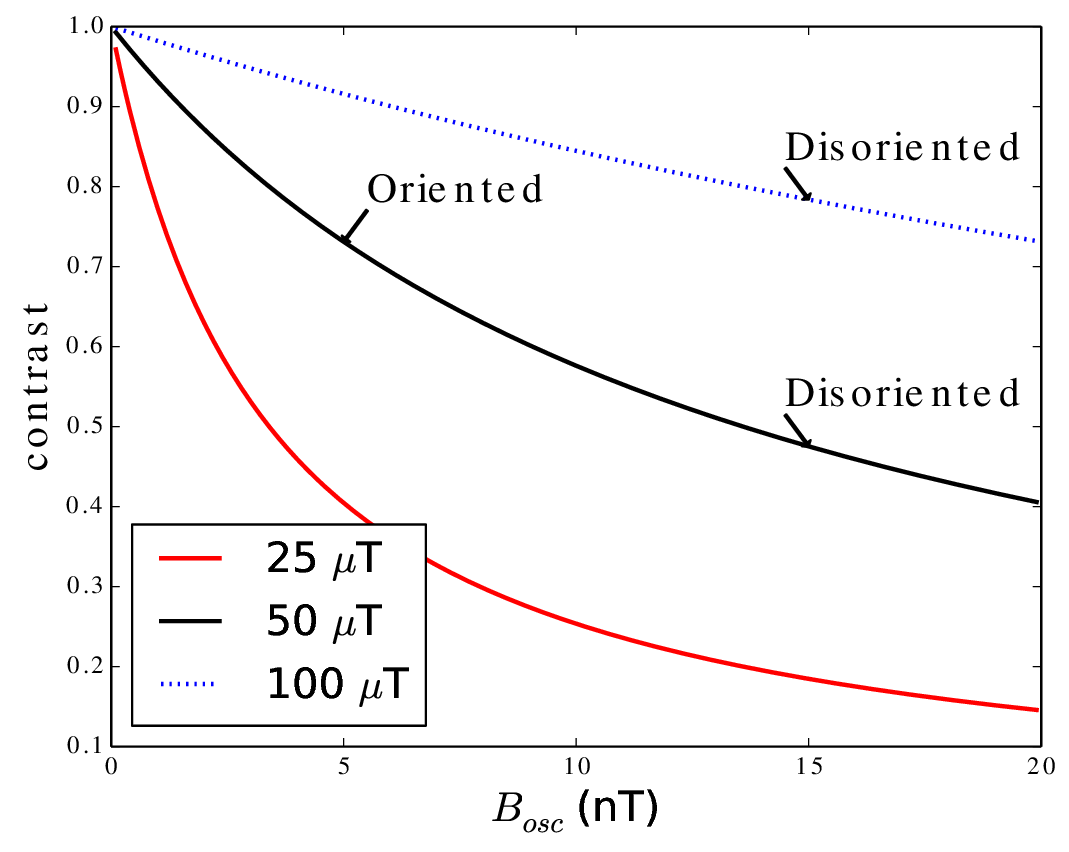}
\end{center}
\caption{(Color online) Contrast between North/South and East/West alignment using
  Eq. \ref{eq:contrast_vs_Bosc}.  Geomagnetic field strength in
  Hamburg, Germany is 47 $\mu$T.  Consistent with
  \cite{ritz2009magnetic}, a rapid loss of contrast occurs as
  $B_{osc}$ increases from 1 to 10 nT.}
\label{fig:contrast_vs_Bosc}
\end{figure}

Similar logic can be used to compare the loss of contrast resulting
from an oscillatory field tuned to the Larmor frequency with that seen
in behavioral experiments.  Here the oscillatory field may flip the
spin of one electron in the radical pair, thereby populating states
with $m_{s}=\pm1$.  As the populations of $\Ket{t^{+}}$, $\Ket{t^{-}}$
and $\Ket{t}$ equilibrate rapidly, the final triplet populations will
be indistinguishable from those produced by the compass mechanism.  As
derived in the appendix, the rate of
such spin flips is $\Omega=\frac{|B_{\text{osc}}|}{2\sqrt{2}}$.
Adding this rate to $R_{ts}(\theta)$ and setting $\Delta \Gamma=0$
yields a new equation for the contrast
\begin{equation}
\text{contrast}=\frac{B^{2} \Delta \mu^{2}}{B^{2} \Delta
  \mu^{2}+|B_{\text{osc}}|\bar{\Gamma}/\sqrt{2}},
\label{eq:contrast_vs_Bosc}
\end{equation}
which is plotted as a function of $B_{\text{osc}}$ in Figure
\ref{fig:contrast_vs_Bosc}.  Consistent with \cite{ritz2004resonance}, Figure
\ref{fig:contrast_vs_Bosc} shows a rapid loss of contrast as
$B_{\text{osc}}$ grows from 1 to 10 nT -- precisely the range in which
experiment shows a crossover from oriented to disoriented behavior.
Some inconsistency with experiment can be seen if the static field
strength is doubled -- while experiment shows disoriented behavior for
$(B,B_{\text{osc}})$=(100 $\mu$T, 15 nT) and oriented behavior for (50
$\mu$T, 5 nT), Figure \ref{fig:contrast_vs_Bosc} shows higher contrast
for the first case than for the second.


When the static field is doubled in the absence of an oscillatory
field, behavioral experiments \cite{wiltschko2006avian} show temporary
disorientation lasting less than an hour, indicating that the
biological signal is affected by the field strength, but the
ability to orient is not.  Here the contrast in Eq. \ref{eq:contrast} is
unaffected by the change in field strength, while the visibility
$R_{ts}(0)-R_{ts}(\pi/2)$ depends on the ratio of $B$ to $k_{s}$.  For
a migratory bird, which is exposed to a range of field strengths, it
may thus be advantageous to have some means of controlling $k_{s}$, so
that the same receptor could give usable visibility at a variety of
field strengths.


The avian compass described in this paper represents a unique example of a
quantum mechanical process which not only survives but is actually sustained by
interaction with a surrounding bath.  Through use of a radical pair, it is
similar to a wide range of biological processes affected by a magnetic field,
including processes as significant as ATP synthesis and DNA replication by
polymerases \cite{buchachenko2014magnetic}.  Although the precise identity of
the receptor or receptors involved in the avian compass remains unknown, simple
geometrical assumptions allow information sufficient for numerical comparison
with experiment to be derived from first principles.  The proposed mechanism
requires neither unique properties nor elaborate manipulation of the radical
pair state, and the biologically observable signal is distinctive and easy to
interpret.  The avian compass thus represents a simple model system for the
emerging and still largely unexplored role of quantum mechanics in biological
processes.

\appendix
\section{Appendix: Dephasing Rates}
The decoherence of a spin system due to interactions with a
surrounding spin bath is one of the central theoretical problems
associated with the avian compass.  It is also a longstanding open
problem in its own right \cite{schlosshauer2007decoherence}.  This
appendix section considers the decay of density matrix
components arising due to hyperfine interactions between two spin 1/2
electrons and a surrounding bath of spin 1/2 nuclei.  Dephasing due to
the bath is treated within the Born-Markov approximation -- the spin
state of each nucleus is assumed to be in thermal equilibrium with the
rest of the bath, to bear no memory of the previous states of the
system or the bath, and to cause decoherence in the central spin
system independently of the other nuclei in the bath.  Having found
rates of decay due to individual nuclei, rates due to the bath as a
whole are found by performing a volume integral over all space assuming
a constant density of nuclei per unit volume.

The hyperfine interaction between a single nucleus and a radical
electron pair is given by
\begin{equation}
  V_{\text{HF}}=\sum_{i,k}(\frac{2\mu_{I}}{I}\frac{1}{|r|^{3}_{ik}}
  )(\vec{I}_{k}\cdot\vec{S}_{i} -3(\vec{I}\cdot\hat{r})(\vec{S}\cdot \hat{r})),
  \label{eq:hyperfine}
\end{equation}
where the $r^{-3}$ dependence of the hyperfine interaction means that
distant electrons interact with effectively distinct reservoirs, while
proximate electrons interact with the same nuclei with comparable
strength.  As selection rules will be important in this derivation,
note that the $I=1/2$ nuclear spin has different angular character
than an $l=1$ magnetic vector field, so that the spin states coupled
in this treatment may differ from effective field approaches.

\subsection{Dephasing in the interaction picture}
The hyperfine interaction between a nucleus and a radical pair contains several
terms with different symmetries, each of which causes the density matrix to
evolve in different ways.  In addition, the system evolves due to the Zeeman
interaction with the magnetic field.  In view of the large size $(8\times8)$
of the matrices involved in these calculations, it is much simpler to treat the decoherence induced by each term separately.

The decay of the density matrix due to the combination of the Zeeman
term and each of the three hyperfine terms will be found in two limits
-- one in which the hyperfine term acts as a perturbation to the
Zeeman term, one in which the Zeeman term acts as a perturbation to
the hyperfine term.  Writing the full Hamiltonian $H=H_{0}+V$ as the
sum of a dominant term $H_{0}$ and a perturbative term $V$, the
evolution of the density matrix can be calculated in the interaction
picture.  For nuclei close to the radical pair, $H_{0}$ is the hyperfine term
and $V$ is the Zeeman term, and vice versa for distant nuclei.

Working in the interaction picture,
\begin{multline}
\frac{\partial}{\partial t} \rho^{I}_{i,k,\epsilon;i\,k',\epsilon'}(t)= \\
-\int_{0}^{\infty} d
\Delta t [V^{I}(t),[V^{I}(t-\Delta
t),\rho^{I}_{i,k,\epsilon;i\,k',\epsilon'}(t-\Delta t)]],
\label{eq:drho_dt}
\end{multline}
where over short times
\begin{equation}
\rho^{I}(t+\Delta t)=e^{-i H_{0} \Delta t}e^{i H \Delta t}\rho^{I}(t)
e^{-i H \Delta t}e^{i H_{0} \Delta t}
\end{equation}
and
\begin{equation}
V(t+\Delta t)=e^{-i H_{0}\Delta t}Ve^{i H_{0}t}.
\end{equation}
Here $i$ indexes electronic states, $k$ nuclear states, and $\epsilon$ the
states of the nucleus's local environment.  Rather than calculate $e^{iHt}$
directly, which would require diagonalizing $H$ anew for every value of
$\alpha/\beta$, the Hamiltonian exponential is approximated by the split
operator method \cite{bandrauk1991improved}
\begin{equation}
  e^{i H \Delta t}
  = e^{i H_{0} \Delta t/2}e^{i V \Delta t} e^{i H_{0}
    \Delta t/2} + \mathbb{O}(\Delta t^{3}),
    \label{eq:splitoperator}
\end{equation}
 so that
\begin{multline}
\rho^{I}(t+\Delta t)\approx \\e^{-i H_{0} \Delta t/2}e^{i V \Delta t}e^{i H_{0} \Delta
  t/2}\rho^{I}(t)
e^{-i H_{0} \Delta t/2}e^{-i V \Delta t}e^{i H_{0} \Delta t/2}.
\label{eq:drhodt_lastform}
\end{multline}

The integrand of Eq. \ref{eq:drho_dt} is now given by the product of a
large number of matrix exponentials multiplying the density matrix, so
that each element of $\dot{\rho}^{I}$ is given by a semi-infinite
integral time integral of a large number of Fourier components.  These
integrals can be evaluated by imposing the Born and Markov
approximations, so that \begin{equation}
  \text{Tr}_{\epsilon=\epsilon'}\rho^{I}_{i,k,\epsilon;i\,k',\epsilon'}(t+\Delta
  t)=
  \rho^{I}_{i,k;i\,k'}(t+\Delta t)\delta(\Delta t)\delta_{k,k'}P_{k},
\label{eq:thermaleq}
\end{equation}
and $\dot{\rho}^{I}_{i,k,;i',k'}=0$ if $k \ne k'$, so that both
$\rho^{I}$ and $\dot{\rho}^{I}$ are diagonal with respect to the
nuclear spin state.  Here the Markov approximation is imposed by
multiplying the integrand by a delta function inside the time
integral, rather than simply replacing $\rho(t+\Delta t)$ with
$\rho(t)$ as in \cite{blum2012density}.  


The time integrals over the various Fourier components can now be
evaluated using a dimensionless integral.  Setting
$\delta(x)=\lim_{\nu\rightarrow \infty}\nu e^{-\nu x}$, where $\nu$ and $x$ are
dimensionless, 
\begin{equation}
\int_{0}^{\infty}dt e^{i \omega t} \delta(t)=\lim_{\nu \rightarrow
  \infty}
\frac{1}{\omega}\int_{0}^{\infty}dx e^{i x}\nu e^{-\nu
  x}=\frac{1}{\omega}.
\label{eq:dimensionless_largeomega}
\end{equation}

In the limit that $e^{i \omega t}$ oscillates slowly relative to the
timescale $\tau$ on which the bath becomes Markovian, the above
integral becomes
\begin{equation}
\int_{0}^{\infty}dt e^{i \omega t} \delta(t)=\lim_{\nu\rightarrow
  \infty} \frac{1}{\kappa} \int_{0}^{\infty}dx e^{i \omega x/\kappa}\nu
e^{-\nu x}=\frac{1}{\kappa},
\label{eq:dimensionless_smallomega}
\end{equation}
where $\kappa=\tau^{-1}$.  Here, Eq. \ref{eq:dimensionless_largeomega}
is used for integrals over oscillating Fourier terms in
Eq. \ref{eq:drho_dt}, while Eq. \ref{eq:dimensionless_smallomega} is
used for integrals over constant terms.

Having found $\dot{\rho}^{I}$ in terms of $\rho^{I}$, the decaying
components of the density matrix and their associated decay rates may
be found by solving an eigenvalue equation.  For the terms involving $\vec{I}\cdot(\vec{S}_{1}\pm \vec{S}_{2})$, tables
\ref{table:Sbar_smallalpha}, \ref{table:Sbar_smallbeta},
\ref{table:deltaS_smallalpha}, and \ref{table:deltaS_smallbeta} give
these rates to second order in $\alpha$ and first order in $\beta$ for
both the symmetric and the antisymmetric hyperfine components, in the
limits that $\alpha<<\beta$ and $\beta<<\alpha$.

\subsection{Matrix forms for the hyperfine and Zeeman interaction}
In order to evaluate Eq. \ref{eq:drho_dt}, it is necessary to have matrix forms
for $H_{0}$ and $V$.  Here it is convenient to decompose the full hyperfine
interaction into three terms, of the form $\vec{I}\cdot(\vec{S}_{1} \pm
\vec{S}_{2})$, and $(\vec{I}\cdot\hat{r})(\vec{\bar{S}}\cdot\hat{r})$.  As each
of these terms have different symmetry, they will cause decay among different
eigencomponents of the density matrix.

\paragraph{Terms involving $\vec{I}\cdot(\vec{S}_{1} \pm \vec{S}_{2}$)}
For the terms involving $\vec{I}\cdot\vec{S}$, rather than treating
the interactions between the nucleus and each electron separately, it
is convenient to reexpress Eq. \ref{eq:hyperfine} in terms of the sum
$\vec{\bar{S}}=\vec{S}_{1}+\vec{S}_{2}$ and the difference $\Delta
\vec{S}=\vec{S}_{1}-\vec{S}_{2}$ of the two spins.  If the distance
$|\vec{R}_{k}|$ between the radical pair and a particular nucleus $k$
is large relative to the spatial extent of the radical pair and the
distance between the two electrons, the hyperfine interaction with
that nucleus can be broken up into two components having different
angular character.  Writing the spatial coordinates of the electrons
as $\vec{\bar{r}}=(\vec{r}_{1}+\vec{r}_{2})/2$ and $\Delta
\vec{r}=(\vec{r}_{1}-\vec{r}_{2})/2$ and assuming that
$|\vec{R}_{k}|>>|\vec{\bar{r}}|$ and $|\vec{R}_{k}|>>|\Delta
\vec{r}|$, the hyperfine interaction with each nucleus $k$ can be
written as the sum of a symmetric term and an antisymmetric term
\begin{equation}
V_{\text{HF}}(\vec{R}_{k},\vec{\bar{r}},\Delta \vec{r})=\sum_{k} V^{(S)}_{\text{HF}}(\vec{R}_{k},\vec{\bar{r}},\Delta \vec{r})+V^{(A)}_{\text{HF}}(\vec{R}_{k},\vec{\bar{r}},\Delta \vec{r})
\end{equation}
where to leading order in the small parameters $|\vec{\bar{r}}|$ and $|\Delta
\vec{r}|$ 
\begin{equation}
V^{(S)}_{\text{HF}}(\vec{R}_{k},\vec{\bar{r}},\Delta \vec{r})=\sum_{k}(\frac{2
\mu_{I}}{I|R_{k}|^{3}}) \vec{I}_{k}\cdot\vec{\bar{S}}
\label{eq:Vs}
\end{equation}
 and
\begin{equation}
  V^{(A)}_{\text{HF}}(\vec{R}_{k},\vec{\bar{r}},\Delta
  \vec{r})=\sum_{k}(\frac{3 \mu_{I}
    \Delta \vec{r}\cdot \vec{R}_{k}}{ I|R_{k}|^{5}}) \vec{I}_{k}\cdot\Delta
  \vec{S}.
\label{eq:Va}
\end{equation}
Note that $\Ket{s}$ and $\Ket{t}$ are eigenstates of
$\vec{\bar{S}}=\vec{S}_{1}+\vec{S}_{2}$, with eigenvalues
$\Ket{\bar{s},\bar{m}}=\Ket{0,0}$ and $\Ket{1,0}$, while states
$\Ket{\uparrow\downarrow}$ and $\Ket{\downarrow\uparrow}$ are eigenstates of
$\Delta \vec{S}=\vec{S}_{1}-\vec{S}_{2}$ with eigenvalues $\Ket{\Delta s,
  \Delta m_{s}}=\Ket{1,\pm1}$.  

Integrating over the spatial component of the wavefunction now leaves
the hyperfine interaction in the form of a spin operator, and the
coefficients of the dot products in Eqs. \ref{eq:Vs} and \ref{eq:Va}
as functions of the nuclear coordinates alone.  Writing
$V_{\text{HF}}^{(S)}(\vec{R}_{k})=\sum_{k}
\alpha(|\vec{R}_{k}|)\vec{I}_{k}\cdot\vec{\bar{S}}
\label{eq:V_HF_S}$
and
$V_{\text{HF}}^{(A)}(\vec{R}_{k})=\sum_{k}
\alpha(|\vec{R}_{k}|,\theta_{k})\vec{I}_{k}\cdot\Delta \vec{S}$, where
$\theta_{k}$ is the angle between $\vec{R}_{k}$ and $\Delta \vec{r}$,
\begin{equation}
\begin{split}
\alpha(|\vec{R}_{k}|)=&\int d^{3}\vec{\bar{r}}\int d^{3}\Delta \vec{r} 
\varphi^{*}(\vec{\bar{r}},\Delta \vec{r}) \frac{2 \mu_{I}}{I|\vec{R}_{k}|^3}
\varphi(\vec{\bar{r}},\Delta \vec{r}) \\
=& \frac{2 \mu_{I}}{I|\vec{R}_{k}|^3}
\end{split}
\label{eq:BS}
\end{equation}
If $\varphi(\vec{\bar{r}},\Delta \vec{r}))=\bar{\varphi}(\vec{\bar{r}})\Delta
F(\Delta r) Y_{l_{o},m_{o}}(\Delta \Omega)$ is separable, with well defined
$l_{o}$ and $m_{o}$,

\begin{equation}
\begin{split}
\alpha(|\vec{R}_{k}|,\theta_{k})=&\int d^{3}\vec{\bar{r}}\int d^{3}\Delta \vec{r} \\
&\varphi^{*} (\vec{\bar{r}},\Delta \vec{r})
 \frac{3 \mu_{I}
    |\Delta \vec{r}||\vec{R}_{k}|\cos(\theta_{k})}{ I|R_{k}|^{5}}
\varphi(\vec{\bar{r}},\Delta \vec{r}) \\
=&\frac{3 \mu_{I} \widetilde{|\Delta \vec{r}|}\widetilde{\cos(\theta_{k})}}
{I |R_{k}|^{4}},
\label{eq:BA}
\end{split}
\end{equation}
where $\widetilde{|\Delta \vec{r}|}=\Bra{\varphi}|\Delta \vec{r}|\Ket{\varphi}$
and $\widetilde{\cos(\theta_{k})}=\Bra{\varphi}\cos(\theta_{k})\Ket{\varphi}$.
Note that the $\widetilde{\cos(\theta_{k})}$ integral introduces a selection rule.
Recalling that $\cos(\theta_{k})$ has angular character $l=1$, with $m_{l}$
dependent upon the orientation, the Wigner-Eckart theorem gives
\begin{equation}
\widetilde{\cos(\theta_{k})}=\Braket{l_{o} || T^{1}|| l_{o}}\Braket{l_{o}
m_{o}1m_{l} | l_{o}m_{o}}
\end{equation}
where $\Braket{l_{o} || T^{1}|| l_{o}}$ is a reduced matrix element and
$\Braket{l_{o}m_{o}1m|l_{o}m_{o}}=\frac{m_{o}}{\sqrt{l_{o}(l_{o}+1)}}$ if
$m_{l}=0$ and $l_{o}\ge 1/2$, but $0$ otherwise.  Setting $m_{o}=0$ eliminates
this term by symmetry.

Having performed these integrals, matrix elements for both components
of the hyperfine interaction have the form $\alpha(\vec{R}_{k})
\vec{I}_{k}\cdot\vec{S}$, where $\vec{S}=\vec{\bar{S}}$ for the
symmetric component and $\vec{S}=\Delta\vec{S}$ for the antisymmetric
component.  Matrix elements of the dot product can be evaluated using
a Clebsch-Gordan expansion
(\cite{woodgate1983elementary} Eq. 9.33)
\begin{equation}
\begin{split}
  \Bra{s'm'_{s} I'm'_{I}}&\vec{I}\cdot\vec{S}\Ket{s m_{s} I
    m_{i}}= \\
&\sum_{J=|s-I|}^{s+I}\sum_{M=-J}^{J}\Braket{s'm'_{s}I'm'_{I}|J M}
  \Braket{s m_{s} I
    m_{i}|JM}\times \\
&\frac{1}{2}(J(J+1)-s(s+1)-I(I+1))\delta_{s,s'}\delta_{I,I'}.
\end{split}
\end{equation}

so that

\begin{equation}
\vec{I}\cdot\vec{S}=
\left(
\begin{array}{cccccccc}
 0 & 0 & 0 & 0 & 0 & 0 & 0 & 0 \\
 0 & 0 & 0 & 0 & 0 & 0 & 0 & 0 \\
 0 & 0 & 0 & 0 & 0 & \frac{1}{\sqrt{2}} & 0 & 0 \\
 0 & 0 & 0 & 0 & 0 & 0 & \frac{1}{\sqrt{2}} & 0 \\
 0 & 0 & 0 & 0 & \frac{1}{2} & 0 & 0 & 0 \\
 0 & 0 & \frac{1}{\sqrt{2}} & 0 & 0 & -\frac{1}{2} & 0 & 0 \\
 0 & 0 & 0 & \frac{1}{\sqrt{2}} & 0 & 0 & \frac{1}{2} & 0 \\
 0 & 0 & 0 & 0 & 0 & 0 & 0 & -\frac{1}{2} \\
\end{array}
\right)
\label{eq:IdotS}
\end{equation}
where the bras and kets represent eigenstates with quantum numbers
$\ket{s,m_{s};m_{I}}$.  Eigenkets and corresponding indices for the
$\vec{S}=\vec{\bar{S}}$ basis are given in Table
\ref{table:Sbar_indices}, and for $S=\Delta \vec{S}$ in Table
\ref{table:deltaS_indices}.  Here the
$\Ket{\bar{s},\bar{m}_{s}}=\Ket{1,\pm1}$ states, although losing
degeneracy with the $m_{s}=0$ subspace in a nonzero magnetic field,
must be included for the sake of second order terms in
Eq. \ref{eq:drho_dt}.

\paragraph{Term involving $(\vec{I}\cdot\hat{r})(\vec{\bar{S}}\cdot\hat{r})$}
The asymmetric term in the hyperfine interaction proportional to
$(\vec{I}\cdot\hat{r})(\vec{\bar{S}}\cdot\hat{r})$ differs
qualitatively from the terms involving $\vec{I}\cdot\vec{\bar{S}}$ and
$\vec{I}\cdot\Delta \vec{S}$ by the presence of a quantization axis
other than the one parallel to the applied magnetic field -- the axis
between the radical pair and the nucleus.  As this paper is concerned with the decoherence between $\ket{s}$ and $\ket{t}$ defined with respect to the magnetic field axis, the operator
\begin{equation}
(\vec{I}\cdot\hat{r})(\vec{\bar{S}}\cdot\hat{r})=
\left(
\begin{array}{cccccccc}
 0 & 0 & 0 & 0 & 0 & 0 & 0 & 0 \\
 0 & 0 & 0 & 0 & 0 & 0 & 0 & 0 \\
 0 & 0 & 0 & 0 & 0 & 0 & 0 & 0 \\
 0 & 0 & 0 & 0 & 0 & 0 & 0 & 0 \\
 0 & 0 & 0 & 0 & \frac{1}{2} & 0 & 0 & 0 \\
 0 & 0 & 0 & 0 & 0 & -\frac{1}{2} & 0 & 0 \\
 0 & 0 & 0 & 0 & 0 & 0 & -\frac{1}{2} & 0 \\
 0 & 0 & 0 & 0 & 0 & 0 & 0 & \frac{1}{2} \\
\end{array}
\right)
\label{eq:IdotSRdotS}
\end{equation}
defined with respect to the $\hat{r}$ axis, with state numbering as defined in table \ref{table:Sbar_indices} is rotated into the $\hat{z}$ axis according to $R(\theta)(\vec{I}\cdot\hat{r})(\vec{\bar{S}}\cdot\hat{r})R(-\theta)$, where
$R(\theta)$ is the outer product of the rotation operators 
\begin{equation}
R_{\text{nuclear}}(\theta)=\left(
\begin{array}{cc}
 \cos (\theta ) & -\sin (\theta ) \\
 \sin (\theta ) & \cos (\theta ) \\
\end{array}
\right)
\end{equation}
and
\begin{equation}
R_{\text{electronic}}(\theta)=\left(
\begin{array}{cccc}
 1 & 0 & 0 & 0 \\
 0 & \cos (\theta ) & -\frac{\sin (\theta )}{\sqrt{2}} & \frac{\sin (\theta
   )}{\sqrt{2}} \\
 0 & \frac{\sin (\theta )}{\sqrt{2}} & \frac{1}{2} (\cos (\theta )+1) &
   \frac{1}{2} (1-\cos (\theta )) \\
 0 & -\frac{\sin (\theta )}{\sqrt{2}} & \frac{1}{2} (1-\cos (\theta )) &
   \frac{1}{2} (\cos (\theta )+1) \\
\end{array}
\right),
\end{equation}
the rotation operators in the nuclear and electronic bases, respectively.

\paragraph{The Zeeman interaction}
In addition to the hyperfine interaction, the system will evolve due
to the influence of the Zeeman Hamiltonian
$H_{z}=\beta(\vec{S}_{1}+\vec{S}_{2})$, given by
\begin{equation}
H_{z}=\beta*\left(
\begin{array}{cccccccc}
 0 & 0 & 0 & 0 & 0 & 0 & 0 & 0 \\
 0 & 0 & 0 & 0 & 0 & 0 & 0 & 0 \\
 0 & 0 & 0 & 0 & 0 & 0 & 0 & 0 \\
 0 & 0 & 0 & 0 & 0 & 0 & 0 & 0 \\
 0 & 0 & 0 & 0 & 1  & 0 & 0 & 0 \\
 0 & 0 & 0 & 0 & 0 & 1  & 0 & 0 \\
 0 & 0 & 0 & 0 & 0 & 0 & -1  & 0 \\
 0 & 0 & 0 & 0 & 0 & 0 & 0 & -1  \\
\end{array}
\right)
\label{eq:zeeman1}
\end{equation}
in the basis diagonalizing $\vec{\bar{S}}$, and
\begin{equation}
H_{z}=\beta*\left(
\begin{array}{cccccccc}
 0 & 0 & 1 & 0 & 0 & 0 & 0 & 0 \\
 0 & 0 & 0 & 1 & 0 & 0 & 0 & 0 \\
 1 & 0 & 0 & 0 & 0 & 0 & 0 & 0 \\
 0 & 1 & 0 & 0 & 0 & 0 & 0 & 0 \\
 0 & 0 & 0 & 0 & 0 & 0 & 0 & 0 \\
 0 & 0 & 0 & 0 & 0 & 0 & 0 & 0 \\
 0 & 0 & 0 & 0 & 0 & 0 & 0 & 0 \\
 0 & 0 & 0 & 0 & 0 & 0 & 0 & 0 \\
\end{array}
\right)
\label{eq:zeeman2}
\end{equation}
in the basis diagonalizing $\Delta \vec{S}$.  Here the Zeeman terms
involving the nuclear magneton, smaller than the Bohr magneton by a
factor of $m_{e}/m_{n}$, have been omitted.

\subsection{Integrated rates of dephasing}
Eigencomponents and rates of dephasing can now be found by substituting the
matrices found in the previous section into Eq. \ref{eq:drho_dt} using Eqs.
\ref{eq:splitoperator}, \ref{eq:thermaleq}, \ref{eq:dimensionless_largeomega},
and \ref{eq:dimensionless_smallomega}.  The
resulting rates are summarized in Tables \ref{table:Sbar_smallalpha},
\ref{table:Sbar_smallbeta}, \ref{table:deltaS_smallalpha},  and
\ref{table:deltaS_smallbeta} for hyperfine interactions of the form
$\vec{I}\cdot(\vec{S}_{1} \pm \vec{S}_{2})$.  Because the strength of the
hyperfine interaction varies with the position of the nucleus within the bath,
these tables give results in terms of $\alpha$, the coefficient of the
$\vec{I}\cdot\vec{S}$ in  Eq. \ref{eq:IdotS} and $\beta$, the coefficient of
the matrix $(\vec{S}_{1}+\vec{S}_{2})$ in Eqs.  \ref{eq:zeeman1} and
\ref{eq:zeeman2}.

The term involving $(\vec{I}\cdot\hat{r})(\vec{\bar{S}}\cdot\hat{r})$
is more computationally difficult than the terms involving
$\vec{I}\cdot(\vec{S}_{1}\pm \vec{S}_{2})$ because the decaying
eigencomponents may vary as a function of the angle $\theta$ between
$\hat{r}$ and $\hat{z}$, the direction of the magnetic field.  As this
paper is primarily concerned with the decay of the $\rho_{st}$ density
matrix component, only that component of $\dot{\rho}_{st}$
proportional to $\rho_{st}$ will be presented here.  For the case when the Zeeman term acts as a perturbation to the hyperfine term, this rate is identically zero, independent of $\theta$.  For the case when the hyperfine term acts as a perturbation to the Zeeman term, this rate is given to second order in $\alpha$ and first order in $\beta$ by
\begin{equation}
	\begin{split}
\dot{\rho}_{st}&=\frac{\alpha ^2 \sin ^2(\theta ) \cos ^3(\theta )}{8 \beta
}-\frac{\alpha ^2 \sin (3 \theta ) \sin (\theta ) \cos ^3(\theta )}{8 \beta
}\\&
+\frac{\alpha ^2 \sin ^4(\theta ) \cos (\theta )}{8 \beta }-\frac{\alpha
^2 \sin (3 \theta ) \sin ^3(\theta ) \cos (\theta )}{8 \beta }\\&
-\frac{\alpha
^2 \sin ^2(\theta ) \cos ^3(\theta )}{8 \omega }+\frac{\alpha ^2 \sin (3 \theta
) \sin (\theta ) \cos ^3(\theta )}{8 \omega }\\&
-\frac{1}{4} \alpha  \sin
^4(\theta ) \cos (\theta)+\frac{1}{4} \alpha  \sin (3 \theta ) \sin ^3(\theta )
\cos (\theta).
   \end{split}
\end{equation}
Noting that
\begin{equation}
\int_{0}^{\pi}\dot{\rho}_{st}(\theta) \sin(\theta) d\theta=0,
\end{equation}
it can be seen that the asymmetric component of the hyperfine
interaction does not contribute to the decay of the singlet-triplet
coherence after integrating over the volume occupied by the bath.

\begin{table}
$\left. \begin{array}{|c|c|c|}
    \hline
  \text{Index} & \Ket{\bar{s}, \bar{m}_{s}; m_{I}} & \Ket{m_{s1}m_{s2};m_{I}} \\ \hline
  1 & \ket{0,0;\uparrow} & (\Ket{\uparrow\downarrow;\uparrow}-\Ket{\downarrow\uparrow;\uparrow})/\sqrt{2} \\
  2 & \ket{0,0;\downarrow} & (\Ket{\uparrow\downarrow;\downarrow}-\Ket{\downarrow\uparrow;\downarrow})/\sqrt{2} \\
  3 & \ket{1,0;\uparrow} & (\Ket{\uparrow\downarrow;\uparrow}+\Ket{\downarrow\uparrow;\uparrow})/\sqrt{2} \\
  4 & \ket{1,0;\downarrow} & (\Ket{\uparrow\downarrow;\downarrow}+\Ket{\downarrow\uparrow;\downarrow})/\sqrt{2} \\
  5 & \ket{1,1;\uparrow} & \Ket{\uparrow\uparrow;\uparrow} \\
  6 & \ket{1,1;\downarrow} & \Ket{\uparrow\uparrow;\downarrow} \\
  7 & \ket{1,-1;\uparrow} & \Ket{\downarrow\downarrow; \uparrow} \\
  8 & \ket{1,-1;\downarrow} & \Ket{\downarrow\downarrow; \downarrow} \\ 
\hline
\end{array} \right.$
\caption{Indices for eigenstates of $\Ket{\bar{s},\bar{m}_{s};m_{I}}$
  and the corresponding kets in the $\Ket{m_{s1}m_{s2};m_{I}}$ basis.}
\label{table:Sbar_indices}
\end{table}

\begin{table}
$\left. \begin{array}{|c|c|c|}
    \hline
  \text{Index} & \Ket{\Delta s, \Delta m_{s}; m_{I}} & \Ket{m_{s1}m_{s2};m_{I}} \\ \hline
  1 & \ket{0,0;\uparrow} &  (\Ket{\uparrow\uparrow;\uparrow}-\Ket{\downarrow\downarrow;\uparrow})/\sqrt{2} \\
  2 & \ket{0,0;\downarrow} & (\Ket{\uparrow\uparrow;\downarrow}-\Ket{\downarrow\downarrow;\downarrow})/\sqrt{2}  \\
  3 & \ket{1,0;\uparrow} &  (\Ket{\uparrow\uparrow;\uparrow}+\Ket{\downarrow\downarrow;\uparrow})/\sqrt{2} \\
  4 & \ket{1,0;\downarrow} & (\Ket{\uparrow\uparrow;\downarrow}+\Ket{\downarrow\downarrow;\downarrow})/\sqrt{2}  \\
  5 & \ket{1,1;\uparrow} & \Ket{\uparrow\downarrow;\uparrow}  \\
  6 & \ket{1,1;\downarrow} &  \Ket{\uparrow\downarrow;\downarrow} \\
  7 & \ket{1,-1;\uparrow} &  \Ket{\downarrow\uparrow;\uparrow} \\
  8 & \ket{1,-1;\downarrow} &  \Ket{\downarrow\uparrow;\downarrow} \\
\hline
\end{array} \right.$
\caption{Indices for eigenstates of $\Ket{\Delta s,\Delta
    m_{s};m_{I}}$ and the corresponding kets in the
  $\Ket{m_{s1}m_{s2};m_{I}}$ basis.}
\label{table:deltaS_indices}
\end{table}

\begin{table}
$\left.
\begin{array}{|c|c|}
\hline
\text{Decay Rate} & \text{Component} \\ \hline
 -\frac{\alpha }{2} & \left.
\begin{array}{c}
 \rho_{82} \\
 \rho_{46}+\rho_{73} \\
 \rho_{37}+\rho_{64} \\
 \rho_{51} \\
 \rho_{28} \\
 \rho_{15} \\
\end{array}
\right. \\
\hline
 \frac{\alpha ^2}{9 \beta }-\frac{\alpha }{3} & \left.
\begin{array}{c}
 \rho_{71} \\
 \rho_{62} \\
 \rho_{26} \\
 \rho_{17} \\
\end{array}
\right. \\
\hline
 -\frac{20 \alpha ^2}{9 \beta } & \left.
\begin{array}{c}
 \rho_{77}-\rho_{44} \\
 \rho_{66}-\rho_{33} \\
\end{array}
\right. \\
\hline
 -\frac{5 \alpha ^2}{3 \beta } & \left.
\begin{array}{c}
 \rho_{86} \\
 \rho_{75} \\
 \rho_{68} \\
 \rho_{57} \\
 \rho_{42} \\
 \rho_{31} \\
 \rho_{24} \\
 \rho_{13} \\
\end{array}
\right. \\
\hline
 -\frac{\alpha ^2}{3 \beta }-\frac{\alpha ^2}{6 \kappa }-\frac{\alpha }{18} &
   \left.
\begin{array}{c}
 \rho_{53} \\
 \rho_{48} \\
 \rho_{84} \\
 \rho_{35} \\
\end{array}
\right. \\
\hline
 -\frac{40 \alpha ^2}{27 \beta }-\frac{2 \alpha }{9} & \left.
\begin{array}{c}
 \rho_{73}-\rho_{46} \\
 \rho_{64}-\rho_{37} \\
\end{array}
\right. \\
\hline
\end{array}
\right.$
\caption{Decaying density matrix components and associated decay rates
  resulting from Zeeman interaction $V=H_{z}=\beta(S_{z1}+S_{z2})$ and
  hyperfine interaction $V=V_{HF}=\alpha(\vec{I}\cdot\vec{\bar{S}})$
  in the limit that $\alpha<<\beta$.  Indices are numbered according
  to Table \ref{table:Sbar_indices}.  Negative rates correspond
  to decay.}
\label{table:Sbar_smallalpha}
\end{table}

\begin{table}
$\left.
\begin{array}{|c|c|}
\hline
\text{Decay Rate} & \text{Component} \\ \hline
 -\frac{\beta }{2} & \left.
\begin{array}{c}
 \rho_{82} \\
 \rho_{51} \\
 \rho_{28} \\
 \rho_{15} \\
\end{array}
\right. \\ \hline
 -\frac{4 \beta ^2}{27 \alpha }-\frac{10 \beta }{9} & \left.
\begin{array}{c}
 \rho_{86} \\
 \rho_{75} \\
 \rho_{68} \\
 \rho_{57} \\
\end{array}
\right. \\ \hline
 -\frac{\beta ^2}{162 \kappa }-\frac{37 \beta }{81} & \left.
\begin{array}{c}
 \rho_{73}-\rho_{46} \\
 \rho_{64}-\rho_{37} \\
\end{array}
\right. \\ \hline
 -\frac{\beta ^2}{18 \kappa }-\frac{\beta }{3} & \left.
\begin{array}{c}
 \rho_{73}+\rho_{46} \left(4-\frac{\beta }{\beta -2 \kappa }\right) \\ 
 \rho_{64}+\rho_{37} \left(4-\frac{\beta }{\beta -2 \kappa }\right) \\ 
\end{array}
\right. \\ \hline
 -\frac{2 \beta ^2}{27 \alpha }-\frac{\beta ^2}{18 \kappa }-\frac{\beta }{6} &
   \left.
\begin{array}{c}
 \rho_{84} \\
 \rho_{71} \\
 \rho_{62} \\
 \rho_{53} \\
 \rho_{48} \\
 \rho_{35} \\
 \rho_{26} \\
 \rho_{17} \\
\end{array}
\right. \\ \hline
\end{array}
\right.$
\caption{Decaying density matrix components and associated decay rates
  resulting from hyperfine interaction
  $H_{0}=\alpha(\vec{I}\cdot\vec{\bar{S}})$ and Zeeman interaction
  $V=H_{z}=\beta(S_{z1}+S_{z2})$ in the limit that $\beta<<\alpha$.
  Indices are numbered according to Table \ref{table:Sbar_indices}.
  Negative rates correspond to decay.}
\label{table:Sbar_smallbeta}
\end{table}

\begin{table}
$\left.
\begin{array}{|c|c|}
\hline
\text{Decay Rate} & \text{Component} \\ \hline
 -\frac{5 \alpha }{12} & \left.
\begin{array}{c}
 \rho_{46}+\rho_{73} \\
 \rho_{37}+\rho_{64} \\
\end{array}
\right. \\ \hline
 -\frac{\alpha ^2}{2 \beta } & \left.
\begin{array}{c}
 \rho_{24}+\rho_{42} \\
 \rho_{13}+\rho_{31} \\
\end{array}
\right. \\ \hline
 -\frac{5 \alpha  \kappa }{6 \beta } & \left.
\begin{array}{c}
 \rho_{86} \\
 \rho_{75} \\
 \rho_{68} \\
 \rho_{57} \\
\end{array}
\right. \\ \hline
 -\frac{\alpha ^2}{12 \kappa }-\frac{5 \alpha }{18} & \left.
\begin{array}{c}
 \rho_{82} \\
 \rho_{15} \\
\end{array}
\right. \\ \hline
 -\frac{2 \alpha ^2}{\beta }+\frac{\alpha ^2}{6 \kappa }-\frac{\alpha }{9} &
   \left.
\begin{array}{c}
 \rho_{77}-\rho_{44} \\
 \rho_{66}-\rho_{33} \\
\end{array}
\right. \\ \hline
 -\frac{\alpha ^2}{2 \beta }-\frac{5 \alpha }{12} & \left.
\begin{array}{c}
 \rho_{71}-\rho_{26} \\
 \rho_{62}-\rho_{17} \\
\end{array}
\right. \\ \hline
 -\frac{\alpha ^2}{2 \beta }-\frac{\alpha ^2}{6 \kappa }+\frac{\alpha }{9} &
   \left.
\begin{array}{c}
 \rho_{42}-\rho_{24} \\
 \rho_{31}-\rho_{13} \\
\end{array}
\right. \\ \hline
 -\frac{\alpha ^2}{2 \beta }-\frac{7 \alpha }{18} & \left.
\begin{array}{c}
 \rho_{26}+\rho_{71} \\
 \rho_{17}+\rho_{62} \\
\end{array}
\right. \\ \hline
 -\frac{2 \alpha ^2}{\beta }-\frac{\alpha }{6} & \left.
\begin{array}{c}
 \rho_{73}-\rho_{46} \\
 \rho_{64}-\rho_{37} \\
\end{array}
\right. \\ \hline
 -\frac{\alpha ^2}{2 \beta }-\frac{\alpha ^2}{12 \kappa }-\frac{5 \alpha }{18} &
   \left.
\begin{array}{c}
 \rho_{84} \\
 \rho_{53} \\
 \rho_{48} \\
 \rho_{35} \\
\end{array}
\right. \\ \hline
 \frac{\alpha ^2}{12 \beta }-\frac{\alpha ^2}{12 \kappa }+\frac{\kappa  \alpha
   }{36 \beta }-\frac{\alpha }{36} & \left.
\begin{array}{c}
 \rho_{51} \\
 \rho_{28} \\
\end{array}
\right. \\ \hline
\end{array}
\right.$
\caption{Decaying density matrix components and associated decay rates
  resulting from Zeeman interaction $V=H_{z}=\beta(S_{z1}+S_{z2})$ and
  hyperfine interaction $V=V_{HF}=\alpha(\vec{I}\cdot\Delta\vec{S})$
  in the limit that $\alpha<<\beta$.  Indices are numbered according
  to Table \ref{table:deltaS_indices}.  Negative rates correspond to
  decay.}
\label{table:deltaS_smallalpha}
\end{table}

\begin{table}
$
\left.
\begin{array}{|c|c|}
\hline
\text{Decay Rate} & \text{Component} \\ \hline
 -\frac{20 \beta ^2}{3 \alpha } & \left.
\begin{array}{c}
 \rho_{42}-\rho_{24} \\
 \rho_{31}-\rho_{13} \\
\end{array}
\right. \\ \hline
 -\frac{5 \beta ^2}{3 \alpha } & \left.
\begin{array}{c}
 \rho_{84} \\
 \rho_{82} \\
 \rho_{73} \\
 \rho_{64} \\
 \rho_{53} \\
 \rho_{51} \\
 \rho_{48} \\
 \rho_{46} \\
 \rho_{37} \\
 \rho_{35} \\
 \rho_{28} \\
 \rho_{26} \\
 \rho_{17} \\
 \rho_{15} \\
\end{array}
\right. \\ \hline
 -\frac{5 \beta ^2}{3 \alpha }+\frac{\beta ^2}{9 \kappa }-\frac{\beta }{18} &
   \left.
\begin{array}{c}
 \rho_{26}+\rho_{71} \\
 \rho_{17}+\rho_{62} \\
\end{array}
\right. \\ \hline
 -\frac{20 \beta ^2}{3 \alpha }+\frac{4 \beta ^2}{9 \kappa }-\frac{2 \beta }{9}
   & \left.
\begin{array}{c}
 \rho_{44}-\rho_{22} \\
 \rho_{33}-\rho_{11} \\
\end{array}
\right. \\ \hline
\end{array}
\right.
$
\caption{Decaying density matrix components and associated decay rates
  resulting from hyperfine interaction
  $H_{0}=\alpha(\vec{I}\cdot\Delta\vec{S})$ and Zeeman interaction
  $V=H_{z}=\beta(S_{z1}+S_{z2})$ in the limit that $\beta<<\alpha$.
  Indices are numbered according to Table \ref{table:deltaS_indices}.
  Negative rates correspond to decay.}
\label{table:deltaS_smallbeta}
\end{table}

\paragraph{Volume integral over the bath}
For the purpose of calculating
dynamics of the avian compass, two rates are particularly important, because
they enter into the equations of motion for the $2\times2$ density matrix
describing dynamics within the $\ket{s}$ and $\ket{t}$ subspace.  These two rates
are given by the integrals over all space of $\frac{5 \alpha^{2}}{3 \beta}$,
the rate of decay for coherence terms $\rho_{13}$, $\rho_{31}$, $\rho_{24}$ and
$\rho_{42}$ in Table \ref{table:Sbar_smallalpha}, and $\frac{20\alpha^{2}}{9
\beta}$, the rate of decay for population imbalances $\rho_{77}-\rho_{44}$ and
$\rho_{66}-\rho_{33}$.  As these rates apply when $\alpha<\beta$, the volume
integral will be performed over all space outside a sphere of radius $r_{0}$,
where $r_{0}$ is the radius at which $\alpha=\beta$, where the hyperfine
interaction between a nuclear and an electronic spin has magnitude
equal to the interaction of the electron spin with the static magnetic field.  

As these rates involve a density of nuclei multiplying a volume integral, it is
helpful to parameterize the result in terms of $N_{\text{sphere}}=(N/V)
\frac{4}{3}\pi r_{0}^{3}$, the number of nuclei within a radius $r_{0}$.
Here
$(N/V)$ is the density of nuclei per unit volume.
Substituting $\beta=B \mu_{B}$,
$\alpha=\alpha_{0}r^{-3}$, $\alpha_{0}r_{0}^{-3}=B \mu_{B}$
yields integrated rates of decay
$\bar{\Gamma}=\frac{5}{3}B \mu_{B}N_{\text{sphere}}$ for elements
$\rho_{13}$, $\rho_{31}$, $\rho_{24}$ and $\rho_{42}$ and $\frac{20}{9}B
\mu_{B}N_{\text{sphere}}$ for elements $\rho_{77}-\rho_{44}$ and
$\rho_{66}-\rho_{33}$.  

$r_{0}$ and $N_{\text{sphere}}$ can be found
by recalling that $\alpha_{0}=(2 \mu_{B} \frac{m_{e}}{m_{p}})/I$ and
$I=1/2$, yielding $r_{0}=42$ bohr for a magnetic field of 50 $\mu$T.
Assuming a density of protons equal to that of liquid water yields
$N_{\text{sphere}}=3300$.  Thus, it is apparent that
$\frac{\bar{\Gamma}}{B \mu_{B}}=\frac{5}{3}\cdot3300 >> 1$, so that
the system is strongly overdamped.  The timescale for decay of
population imbalances $\rho_{77}-\rho_{44}$ and
$\rho_{66}-\rho_{33}$ is given by $(\frac{20}{9}B \mu_{B}
N_{\text{sphere}})^{-1}=33$ ps, so that any
population transferred to the $\Ket{s,m_{s}}=\Ket{1,0}$ state will
quickly equilibrate with the populations of states $\Ket{1,\pm 1}$.

\subsection{Rabi oscillation in the limit of strong dephasing}
In \cite{ritz2004resonance}, an oscillatory magnetic field tuned to the Larmor
frequency for an electron in the static geomagnetic field was found to
cause disorientation in European robins.  In the body of the paper,
this was attributed to electrons flipping their spin due to the
oscillatory field, creating an alternate pathway for the formation of
triplet state population from an initial singlet state which does not
depend on the orientation of the compass molecule.  Because the
populations of triplet states $\Ket{t^{+}}$, $\Ket{t^{-}}$ and
$\Ket{t}$ equilibrate very rapidly, the population of triplet
states created in this way will be indistinguishable from those
created by the compass mechanism.

In the absence of dephasing, the rate of spin flips due to Hamiltonian 
$H=\left(
\begin{array}{cc}
 \frac{\omega_{1}}{2} & d \cos(\omega_{2} t) \\
 d \cos (\omega_{2} t) & -\frac{\omega_{1}}{2} \\
\end{array}
\right)$ is the Rabi frequency
$\Omega_{\text{Rabi}}=\sqrt{d^{2}+\Delta^2}$, where
$\Delta=\omega_{2}-\omega_{1}$ is the detuning between the driving
frequency and the spacing between the two energy levels.  As shown
elsewhere in this section, the dynamics of the radical pair density
matrix can be greatly affected by dephasing induced by the surrounding
nuclear spin bath; thus, a brief discussion of Rabi oscillation in the
limit of rapid dephasing is warranted.  Here the effects of dephasing
are given by a Lindblad term $\mathcal{L}[\rho]=-\rho L^{\dag}
L-L^{\dag} L \rho+2 L\rho L^{\dag}$, where 
$L=\left(
\begin{array}{cc}
 1 & 0 \\
 0 & 0 \\
\end{array}
\right)$, so that the density matrix obeys
\begin{equation}
\frac{d}{dt}\rho=-i[H,\rho]+\Gamma\mathcal{L}[\rho].
\end{equation}

Writing $\rho=\left(
\begin{array}{cc}
 \rho_{11} & e^{-i t \omega_{1}} \rho_{12} \\
 e^{i t \omega_{1}} \rho_{21} & \rho_{22} \\
\end{array}
\right)$, equations for the slowly evolving $\rho_{ij}$ can be found
by imposing the rotating wave approximation $e^{i n
  \omega_{1}t}\rightarrow 0$ for $n \ne 0$, so that
\begin{equation}
\left(
\begin{array}{c}
 \dot{\rho}_{11} \\ \dot{\rho}_{12} \\
 \dot{\rho}_{21} \\ \dot{\rho}_{22} \\
\end{array}
\right)=
\left(
\begin{array}{c}
 \frac{1}{2} i d e^{i t \Delta } \rho_{12}-\frac{1}{2} i d e^{-i t \Delta }
   \rho_{21} \\ \frac{1}{2} i d e^{-i t \Delta } \rho_{11}-\frac{1}{2} i d
   e^{-i t \Delta } \rho_{22}-\rho_{12} \Gamma  \\
 -\frac{1}{2} i d e^{i t \Delta } \rho_{11}+\frac{1}{2} i d e^{i t \Delta }
   \rho_{22}-\rho_{21} \Gamma  \\ \frac{1}{2} i d e^{-i t \Delta }
   \rho_{21}-\frac{1}{2} i d e^{i t \Delta } \rho_{12} \\
\end{array}
\right)
\label{eq:drhodt_slowterms}
\end{equation}

A second order differential equation for the density matrix components
can now be found by taking the time derivative of
Eq. \ref{eq:drhodt_slowterms}, then using
Eq. \ref{eq:drhodt_slowterms} to substitute for terms of the form
$\dot{\rho}_{ij}$, yielding
\begin{widetext}
\begin{equation}
\left(
\begin{array}{c}
 \ddot{\rho}_{11} \\ \ddot{\rho}_{12} \\
 \ddot{\rho}_{21} \\ \ddot{\rho}_{22} \\
\end{array}
\right)=
\left(
\begin{array}{c}
 -\frac{\rho_{11} d^2}{2}+\frac{\rho_{22} d^2}{2}-\frac{1}{2} i e^{i t \Delta
   } \rho_{12} \Gamma  d+\frac{1}{2} i e^{-i t \Delta } \rho_{21} \Gamma 
   d-\frac{1}{2} e^{i t \Delta } \rho_{12} \Delta  d-\frac{1}{2} e^{-i t \Delta
   } \rho_{21} \Delta  d \\ -\frac{\rho_{12} d^2}{2}+\frac{1}{2} e^{-2 i t
   \Delta } \rho_{21} d^2-\frac{1}{2} i e^{-i t \Delta } \rho_{11} \Gamma 
   d+\frac{1}{2} i e^{-i t \Delta } \rho_{22} \Gamma  d+\frac{1}{2} e^{-i t
   \Delta } \rho_{11} \Delta  d-\frac{1}{2} e^{-i t \Delta } \rho_{22} \Delta 
   d+\rho_{12} \Gamma ^2 \\
 \frac{1}{2} e^{2 i t \Delta } \rho_{12} d^2-\frac{\rho_{21}
   d^2}{2}+\frac{1}{2} i e^{i t \Delta } \rho_{11} \Gamma  d-\frac{1}{2} i e^{i
   t \Delta } \rho_{22} \Gamma  d+\frac{1}{2} e^{i t \Delta } \rho_{11} \Delta
    d-\frac{1}{2} e^{i t \Delta } \rho_{22} \Delta  d+\rho_{21} \Gamma ^2 \\
   \frac{\rho_{11} d^2}{2}-\frac{\rho_{22} d^2}{2}+\frac{1}{2} i e^{i t \Delta
   } \rho_{12} \Gamma  d-\frac{1}{2} i e^{-i t \Delta } \rho_{21} \Gamma 
   d+\frac{1}{2} e^{i t \Delta } \rho_{12} \Delta  d+\frac{1}{2} e^{-i t \Delta
   } \rho_{21} \Delta  d \\
\end{array}
\right)
\label{eq:drhodt}
\end{equation}
\end{widetext}
In the limit that $\Gamma>>d$, terms of order $\Gamma^{2}$ dominate
Eq. \ref{eq:drhodt}, so that $\rho_{12}$ and $\rho_{21}$ decay as
$e^{-\Gamma t}$.  Substituting $\rho_{12},\rho_{21}\rightarrow 0$ into
Eq. \ref{eq:drhodt} yields a second order differential equation for
the population difference 
\begin{equation}
\ddot{\rho}_{11}-\ddot{\rho}_{22}=-d^{2}(\rho_{11}-\rho_{22}),
\end{equation}
so that the population difference oscillates as 
\begin{equation}
(\rho_{11}(t)-\rho_{22}(t))=A \sin(d t) + B \cos(dt).
\label{eq:drhosqdtsq}
\end{equation}
Note that the rapid decay of $\rho_{12}$ and $\rho_{21}$ cause terms
involving $\Delta$ to vanish from the equation for the population
difference, so that the effects of nonzero detuning are small in the limit of
rapid dephasing.


For a radical pair in an initial singlet state, it is the difference
$\Delta \mu$ between the effective magnetic moments of the two spins
which drives spin flips.  Using $H^{rp}_{0}=1/2 \Delta \mu
\vec{B}\cdot\Delta\vec{S}$ from Eq. \ref{eq:H0rp} in the body of the
paper, the three states with $\Delta s=1$, $\Delta m_{s}=-1,0,1$
are coupled by the oscillatory field according to
$\frac{1}{2}B(t) \Delta \mu \sigma_{x}=\frac{B_{\text{osc}}\Delta
  \mu}{2\sqrt{2}}\cos(\omega_{2} t)\left(
\begin{array}{ccc}
 0 & 1 & 0 \\
 1 & 0 & 1 \\
 0 & 1 & 0 \\
\end{array}
\right)$.  Setting $\Delta \mu=1$ as in the body of the paper yields
$d=\frac{B_{\text{osc}}}{2\sqrt{2}}$ for both of the two Rabi pathways.

\end{document}